\documentclass[11pt]{article}
\usepackage{amsmath}
\usepackage{amsfonts}
\usepackage{amssymb}
\usepackage[pdftex]{color,graphicx}
\usepackage{tikz}
\usetikzlibrary{matrix,arrows,decorations.pathmorphing}
\usepackage{tikz-cd}
\usetikzlibrary{arrows}

\newtheorem{Theorem}{Theorem}

\newtheorem{Cor}{Corollary}

\begin{document}

\def\Z{\mathbb{Z}}
\def\R{\mathbb{R}}
\def\C{\mathbb{C}}
\def\S{\mathbb{S}}
\def\T{\mathbb{T}}
\def\H{\mathbb{H}}

\newcommand{\ket}[1]{|{#1}\rangle}
\newcommand{\bra}[1]{\langle{#1}|}
\newcommand{\states}{\mathcal{S}}

\newcommand{\set}[1]{\ensuremath{ \lbrace #1 \rbrace }}
\newcommand{\Tr}{\text{Tr}}
\newcommand{\Span}[1]{\ensuremath{ \langle #1 \rangle }}
\newcommand{\ttd}{{\tt{d}}}
\newcommand{\sfr}{\mathfrak{s}}

\def\pP{\mathcal{P}}
\def\cC{\mathcal{C}}
\def\oO{\mathcal{O}}
\def\iI{\mathcal{I}}
\def\eE{\mathcal{E}}

\def\ss{\mathfrak{s}}

\title{Putting paradoxes to work: contextuality in measurement-based quantum computation}

\author{Robert Raussendorf\vspace{4mm}\\
{\small{\em{Department of Physics \& Astronomy, University of British Columbia, Vancouver, Canada,}}}\\
{\small{\em{Stewart Blusson Quantum Matter Institute, University of British Columbia, Vancouver, Canada}}}
}

\maketitle

\begin{abstract}We describe a joint cohomological framework for measurement-based quantum computation (MBQC) and the corresponding contextuality proofs. The central object in this framework is an element $[\beta_\Psi]$ in the second cohomology group of the chain complex describing a given MBQC. $[\beta_\Psi]$ contains the function computed therein up to gauge equivalence, and at the same time is a contextuality witness. The present cohomological description only applies to temporally flat MBQCs, and we outline an approach for extending it to the temporally ordered case.
\end{abstract}

\section{Introduction}

Spectators may be sent into infinite loops by Zeno, but Achilles catches up with the turtle anyway. Paradoxes do not spell trouble in the way contradictions do, as a contradiction appears in them only when improper  assumptions are made. The more reasonable these assumptions seem, the brighter shines the paradox. 

Here we investigate the paradox of Kochen and Specker  \cite{KS}, \cite{Bell}, describing a particular property of quantum mechanics by which it is distinguished from classical physics: contextuality \cite{KS}--\cite{ID}. The statement ``quantum mechanics is contextual" means that descriptions of quantum phenomena in terms of classical statistical mechanics---so-called non-contextual hidden variable models (ncHVMs) \cite{EPR},\cite{KS},\cite{Bell}---are in general not viable.  In such models, all observables are assigned pre-existing values which are merely revealed by measurement---in stark contrast with quantum mechanics.\medskip

Each paradox invites us to ask what becomes of the glaring discrepancy once the (in hindsight) improper assumption is excised. For the Kochen-Specker paradox, one such inquiry leads to measurement-based quantum computation (MBQC) \cite{RB01}, a scheme of universal quantum computation driven by measurement. 

We identify the mathematical structures that simultaneously capture the contextuality and the computational output of measurement-based quantum computations. These structures turn out to be  cohomological.  Put in graphical form, we explore the following triangle.
\begin{equation}\label{Triangle}
\parbox{10cm}{\includegraphics[width=10cm]{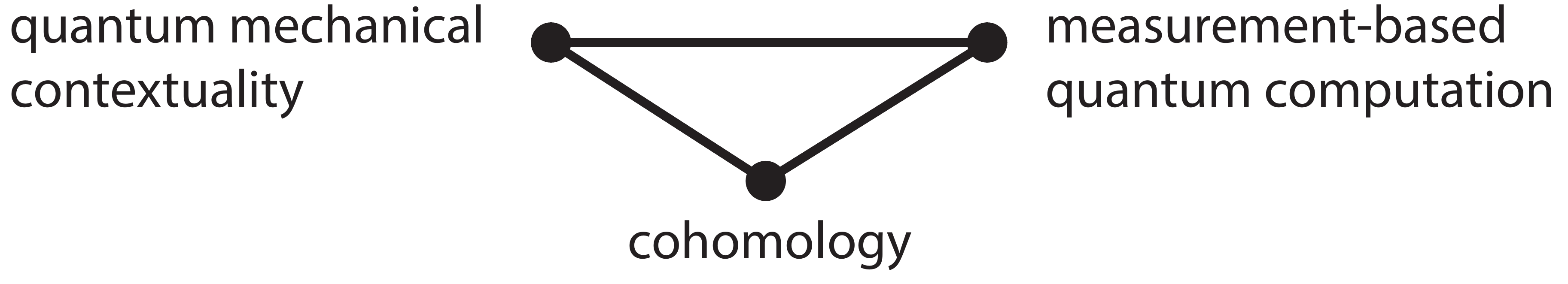}}
\end{equation}
In the first part of this paper, consisting of Sections~\ref{bg} and \ref{CohoW}, we flesh out the above diagram for the simplest case, deterministic temporally flat MBQCs and the corresponding proofs of contextuality. Temporally flat means that these MBQCs have no non-trivial temporal order, which is a restriction. Section~\ref{bg} reviews the necessary background on contextuality and measurement-based quantum computation, and Section~\ref{CohoW} explains how cohomology encapsulates the essence of parity-based contextuality proofs and temporally flat MBQCs. The main results are Theorem~\ref{CPth2} \cite{Coho} and Theorem~\ref{ObeT} \cite{CohoMBQC}, which we restate here.

In the second part of the paper, consisting of Section~\ref{TO}, we work towards removing the assumption of temporal flatness. MBQCs are typically temporally ordered. Even though the measurements driving the computation commute, measurement bases need to be adjusted depending on the outcomes of  other measurements, and this introduces temporal order. This adjustment is necessary to prevent the randomness inherent in quantum measurement from creeping into the logical processing. 

While we do not yet tackle temporally ordered MBQCs, we demonstrate that a known contextuality proof exhibiting temporal ordering of measurements, the so-called ``iffy'' proof \cite{Exa}, can be described by the {\em{same}} cohomological formalism that is used for the temporally flat case. We conjecture that this strategy might also work for general MBQCs. 

Section~\ref{Concl} is the conclusion, and Section~\ref{TL} covers some stations of the author's own journey through the world of quantum computation and paradox.

\section{Background}\label{bg}

\subsection{Contextuality}\label{Crev}

We assume that the reader is familiar with the concept of contextuality \cite{KS},\cite{Bell}; see \cite{Merm} for a review. To provide a short summary, contextuality of quantum mechanics signifies that, in general, quantum mechanical phenomena cannot be described by so-called non-contextual hidden variable models (ncHVMs) \cite{EPR}. In an ncHVM, observable quantities have predetermined value assignments; i.e., each observable possesses a value, and those values are merely revealed upon measurement. The statistical character of measurement in quantum mechanics is then sought to be reproduced by a probability distribution over the value assignments. For certain sets of measurements no such probability distribution exists. If that's the case, then the physical setting at hand is contextual.

In this paper, we assume that in each value assignment $\lambda$ is deterministic; i.e., the value $\lambda(A)$ assigned to each observable $A$ is an eigenvalue of that observable, in accordance with the Dirac projection postulate. More general constructs are conceivable; for example the value assignments may themselves be probability distributions over eigenvalues \cite{Spekk}; however, we do not consider such generalizations here. We remark that deterministic ncHVMs are equivalent to factorizable probabilistic ones \cite{AB}; also see \cite{Fine}. \smallskip 

The Kochen-Specker (KS) theorem~\cite{KS} says that in Hilbert spaces of dimension 3 and higher, it is impossible to assign all quantum-mechanical observables deterministic non-contextual values in a consistent fashion. A very simple proof of the KS theorem, in dimension 4 and up, is provided by Mermin's square \cite{Merm}. It is the simplest parity proof of contextuality, where the assumption of existence of a consistent non-contextual value assignment $\lambda$ leads to a system of mod 2--linear equations with an internal inconsistency. As we will discuss below, the connection between contextuality and MBQC runs through the parity proofs.

For MBQC we employ state-dependent contextuality. In it, consistent value assignments $\lambda$ do exist, but no probability distribution over them can explain the measurement statistics for the quantum state in question. The reason that value assignments suddenly become possible does not contradict the KS theorem; we merely have shrunk the set of observables considered. Already the original proof \cite{KS} of the KS theorem and the simpler proof via Mermin's square use a finite number of observables picked from a priori infinite sets; and in the application to MBQC we simply reduce those sets further.\smallskip

The key example for the connection between contextuality and MBQC is the state-dependent version of Mermin's star \cite{Merm}, as was observed in \cite{AB}. Consider the eight-dimensional Hilbert space of 3 qubits, a specific state in it, the Greenberger-Horne-Zeilinger (GHZ) state \cite{GHZ},
\begin{equation}\label{GHZ}
|\text{GHZ}\rangle = \frac{|000\rangle + |111\rangle}{\sqrt{2}},
\end{equation}
and furthermore the six local Pauli observables $X_i$, $Y_i$, $i=1,..,3$. The state-dependent contextuality question is whether those six local observables can be assigned values $\lambda(\cdot) = \pm 1$ in such a way that the measurement statistics for the four non-local Pauli observables 
$X_1X_2X_3$, $X_1Y_2Y_3$, $Y_1X_2Y_3$, $Y_1Y_2X_3$
is reproduced.

The GHZ state is a simultaneous eigenstate of these observables,
$$
X_1X_2X_3 \, |\text{GHZ}\rangle =  -X_1Y_2Y_3\,  |\text{GHZ}\rangle =-Y_1X_2Y_3\,  |\text{GHZ}\rangle= -Y_1Y_2X_3 \,  |\text{GHZ}\rangle  = |\text{GHZ}\rangle.
$$ 
The measurement outcomes for the four non-local observables are deterministic and equal to $\pm 1$. 

Now note that these non-local observables are products of the local ones $X_i$, $Y_i$, namely $X_1X_2X_3 = (X_1) (X_2) (X_3)$, $X_1Y_2Y_3 = (X_1) (Y_2) (Y_3)$, etc. Assuming an ncHVM value assignment $\lambda$ for the local observables, the above operator constraints translate into constraints on the assigned values $\lambda(\cdot)$, namely $\lambda(X_1)\lambda(X_2) \lambda(X_3)=+1$, $\lambda(X_1)\lambda(Y_2) \lambda(Y_3)=-1$, and two more of the same kind. It is useful to write the value assignments $\lambda$ in the form $\lambda(\cdot)=(-1)^{s(\cdot)}$. In terms of the binary variables $s$, the four constraints read
\begin{equation}\label{sdMS}
\begin{array}{rcl}
s(X_1) + s(X_2) + s(X_3) \mod 2 &=& 0,\\
s(X_1) + s(Y_2) + s(Y_3) \mod 2 &=& 1,\\
s(Y_1) + s(X_2) + s(Y_3) \mod 2 &=& 1,\\
s(Y_1) + s(Y_2) + s(X_3) \mod 2 &=& 1.\\
\end{array}
\end{equation}
Adding those four equations mod 2 reveals a contradiction $0=1$, hence no value assignment $s$ (equivalently $\lambda$) for the six local observables reproduces the measurement statistics of the GHZ state. The state-dependent Mermin star is thus contextual. We will return to Eq.~(\ref{sdMS}) throughout, as it relates to the simplest example of a contextual MBQC \cite{AB}.
\medskip

In preparation for the subsequent discussion we review one further concept,  the contextual fraction \cite{ABsheaf}. To define it, consider an empirical model $e$, i.e., a collection of probability distributions over measurement contexts, and split it into a contextual part $e^C$ and a non-contextual part $e^{NC}$,
\begin{equation}
e=\tau e^{NC} + (1-\tau) e^C,\; 0 \leq \tau \leq 1.
\end{equation}
The maximum possible value of $\tau$  is called the non-contextual fraction ${\sf{NCF}}(e)$ of the model $e$,
\begin{equation}
{\sf{NCF}}(e) := \max_{e^{NC}} \tau.
\end{equation}
The contextual fraction ${\sf{CF}}(e)$ is then the probability weight of the contextual part $e^{C}$,
\begin{equation}
{\sf{CF}}(e):=1-{\sf{NCF}}(e).
\end{equation}
It is a measure of the ``amount'' of contextuality contained in a given physical setup.

\subsection{Measurement-based quantum computation}\label{MBQCrev}

Again, we assume that the reader is familiar with the concept of measurement-based quantum computation, a.k.a. the one way quantum computer \cite{RB01}. Here we provide only a very short summary, and then expand on one technical aspect that is of particular relevance for the connection with contextuality--the classical side processing. For a review of MBQC see e.g. \cite{RW12}. 

In MBQC, the process of quantum computation is driven by local measurement on an initially entangled quantum state; no unitary evolution takes place. Further, the initial quantum state, for example a 2D cluster state, does not carry any information about the algorithm to be implemented---it is universal. All algorithm-relevant information is inputted to that quantum state, processed and read out by the local measurements.

In quantum mechanics, the basis of a measurement can be freely chosen but the measurement outcome is typically random; and this of course affects MBQC. There, the choice of measurement bases encodes the quantum algorithm to be implemented, and the measurement record encodes the computational output. In MBQC every individual measurement outcome is in fact completely random, and meaningful information is contained only in correlations of measurement outcomes. As it turns out, these computationally relevant correlations have a simple structure. To extract them from the measurement record, every MBQC runs a classical side processing.

The need for classical side processing in MBQC also arises in a second place: measurement bases must be adapted according to previously obtained measurement outcomes, in order to prevent the randomness of quantum measurement from creeping into the logical processing.

We confine our attention to the original MBQC scheme on cluster states \cite{RB01}, which we will henceforth call $l2$-MBQC. There are other MBQC schemes, for example using AKLT states as computational resources, in which the side processing is more involved.

In $l2$-MBQC, for each measurement $i$ there are two possible choices for the measured observable $O_i[q_i]$, depending on a binary number $q_i$. The eigenvalues of these observables are constrained to be $\pm 1$. Furthermore, both the bitwise output $\textbf{o}=(o_1,o_2..,o_k)$ and the choice of measurement bases, $\textbf{q}=(q_1,q_2,..,q_N)$ are functions of the measurement outcomes $\textbf{s}=(s_1,s_2,..,s_N)$. In addition, $\textbf{q}$ is also a function of the classical input $\textbf{i}=(i_1,i_2,..,i_m)$. These functional relations are all mod 2 linear,
\begin{subequations}\label{CCR}
\begin{align}\label{CCR_out}
\textbf{o}&=Z\textbf{s} \mod 2,\\ 
\label{CCR_in}
\textbf{q} &=T\textbf{s}+S\textbf{i} \mod 2.
\end{align}
\end{subequations}
Therein, the binary matrix $T$ encodes the temporal order in a given MBQC. If $T_{ij}=1$ then the basis for the measurement $i$ depends on the outcome of measurement $j$, hence the measurement $j$ must be executed before the measurement $i$. We remark that Eqs.~(\ref{CCR}) have been discussed with additional constant offset vectors on the r.h.s. \cite{TO_sym}, but we don't need that level of generality here.

\subsection{Links between contextuality and MBQC}\label{Link}

The basic result relating MBQC to contextuality is the following.

\begin{Theorem}[\cite{RR13}]\label{NLPCrel}
Be ${\cal{M}}$ an $l2$-MBQC evaluating a function $o:(\mathbb{Z}_2)^m \longrightarrow \mathbb{Z}_2$. Then, ${\cal{M}}$ is contextual if it succeeds with an average probability $p_S>1-d_H(o)/2^m$, where $d_H(o)$ is the Hamming distance of $o$ from the closest linear function.
\end{Theorem}
That is, if the function evaluated by the $l2$-MBQC is non-linear---hence outside what the classical side processing can compute by itself---then the assumption of non-contextuality puts a limit on the reachable probability of success. The reliability of the MBQC can be improved beyond this threshold only in the presence of contextuality. The more nonlinear the computed function (in terms of the Hamming distance $d_H(o)$), the lower the threshold.
The lowest contextuality thresholds are reached for bent functions. For $m$ even and $o$ bent, it holds that $d_H(o) = 2^{m-1} - 2^{m/2-1}$ \cite{MWS}, and therefore the contextuality threshold for the average success probability $p_S$ approaches $1/2$ for large $m$. An MBQC can thus be contextual even though its output is very close to completely random.\medskip
 
In particular when comparing the above Theorem~\ref{NLPCrel} to structurally similar theorems on the role of entanglement in MBQC \cite{MVdN}, we observe that the above only provides a binary ``can do vs. cannot do'' separation. According to the theorem, in the presence of contextuality a success probability of unity is a priori possible, but without it the stated bound applies. Yet it is intuitively clear that the reachable success probability of function evaluation in MBQC should depend on the ``amount'' of contextuality present. In this regard, we note the following refinement of Theorem~\ref{NLPCrel}, invoking the contextual fraction.
\begin{Theorem}[\cite{CF}]\label{T1}
Let $f: (\mathbb{Z}_2)^m \longrightarrow \mathbb{Z}_2$ be a Boolean function, and $\H(f,{\cal{L}})$ its Hamming distance to the closest linear function. For each l2-MBQC with contextual fraction ${\sf{CF}}(\rho)$ that computes $f$ with average success probability $\overline{p}_S$ over all $2^m$ possible inputs it holds that 
\begin{equation}\label{pSbd}
\overline{p}_S\leq 1- \frac{(1-{\sf{CF}}(\rho))\, \H(f,{\cal{L}})}{2^m}.
\end{equation}
\end{Theorem}
Thus, the larger the contextual fraction, the larger the achievable success probability for function evaluation through MBQC. If the contextual fraction of the resource state becomes unity, then the theorem puts no non-trivial bound on the success probability of the corresponding $l2$-MBQC.

If, on the other hand, the contextual fraction of the resource state becomes zero, i.e., when the resource state can be described by a non-contextual hidden variable model, the threshold in success probability reduces to that of Theorem~\ref{NLPCrel}. Theorem~\ref{T1} interpolates between those two limiting cases.\medskip

One important aspect of the MBQC--contextuality relationship is revealed only by the proof of Theorem~\ref{NLPCrel}, but not by the statement of the theorem itself. Namely, the contextuality of MBQC is intimately related to the classical side processing Eq.~(\ref{CCR}). Rather than replicating the proof from \cite{RR13}, here we illustrate the idea through the example of Anders and Browne's GHZ-MBQC \cite{AB}, related to Mermin's star. We will return to this example throughout.\smallskip

{\em{Example (GHZ-MBQC).}} In this scenario, the resource state is a Greenberger-Horne-Zeilinger state of Eq.~(\ref{GHZ}), and the local measurable observables $O_i[q_i]$, depending on a binary number $q_i$, are $O_i[0]=X_i, \; O_i[1]=Y_i$, for $i=1,..,3$. These are precisely the ingredients of the state-dependent version of Mermin's star, as we discussed in Section~\ref{Crev}. As before, the measurement outcomes $s_i\in \mathbb{Z}_2$ are related to the measured eigenvalues $\lambda_i = \pm 1$ of the respective local Pauli observables via $\lambda_i=(-1)^{s_i}$. There are two bits $y,z$ of input and one bit $o$ of output, and the computed function is an OR-gate, $o= y \vee z$. 

The required linear classical side processing is as follows. 
\begin{subequations}\label{CCRghz}
\begin{align}
\label{CCR_inGHZ}
q_1 = y,\, q_2 = z,\, q_3 = y+z \mod 2,\\
\label{CCR_outGHZ}
o= s_1+s_2+s_3 \mod 2.
\end{align}
\end{subequations}
The two input bits $y$ and $z$ determine the choices $q_i$ of measured observables through Eq.~(\ref{CCR_inGHZ}), and then the corresponding binary measurement outcomes $s_1, s_2, s_3$ determine the outputted value of the function, $o(y,z)$.

Let's verify that the output is the intended OR function. First, consider $y=z=0$. Thus, by Eq.~(\ref{CCR_inGHZ}), $q_1=q_2=q_3=0$, and all three locally measured observables are of $X$-type. While the outcomes $s_1,s_2,s_3$ are individually random, they are correlated since the product of the corresponding observables $X_i$ is the stabilizer of the GHZ state, $X_1X_2X_3|\text{GHZ}\rangle = |\text{GHZ}\rangle$. Therefore, $s_1+s_2+s_3\mod 2 = 0$. Hence, with Eq.~(\ref{CCR_outGHZ}), $o(0,0)=0$ as required for the OR-gate.

We consider one more input combination, $y=0$ and $z=1$. Then, with Eq.~(\ref{CCR_inGHZ}), $q_1=0$ and $q_2=q_3=1$. Hence $X_1$, $Y_2$ and $Y_3$ are measured. Because of the stabilizer relation $X_1Y_2Y_3|\text{GHZ}\rangle = - |\text{GHZ}\rangle$, the three measurement outcomes $s_1,s_2,s_3$ satisfy $s_1+s_2+s_3 \mod 2 =1$. With Eq.~(\ref{CCR_outGHZ}), $o(0,1)=1$ as required. The discussion of the other two inputs is analogous.\smallskip

The OR-gate is a very simple function; yet it is of consequence for the above computational setting. Every MBQC requires a classical control computer, to enact the classical side processing of Eq.~(\ref{CCRghz}). This control computer is constrained to performing addition mod 2, and it is therefore not classically computationally universal. The OR-gate is a non-linear Boolean function. By adding it to the available operations, the extremely limited classical control computer is boosted to classical computational universality \cite{AB}.

To understand the connection between contextuality and MBQC classical processing relations, we state Eq.~(\ref{CCR_outGHZ}) separately for all four input values.
\begin{equation}\label{GHZmbqc1}
\begin{array}{rrcr}
\textbf{input:} \; (0,0)& \hspace*{4mm}\textbf{output:} \; 0&=& s(X_1)+s(X_2)+s(X_3)\\
(0,1)& 1&=& s(X_1)+s(Y_2)+s(Y_3)\\
(1,0)& 1&=& s(Y_1)+s(X_2)+s(Y_3)\\
(1,1)& 1&=& s(Y_1)+s(Y_2)+s(X_3)
\end{array}
\end{equation}
Note the striking resemblance of Eq.~(\ref{GHZmbqc1}) to the earlier Eq.~(\ref{sdMS}). The only difference is that Eq.~(\ref{GHZmbqc1}) refers to quantum mechanical measurement record, one context at a time, whereas Eq.~(\ref{sdMS}) refers to a noncontextual value assignment in an ncHVM, applying to all contexts simultaneously. Thus, if we assume an ncHVM then we obtain a contradiction; and if we do not assume it then the same equations describe a computation.

This dichotomy exists not only for the GHZ-scenario discussed here, but indeed for all MBQCs satisfying the classical processing relations Eq.~(\ref{CCR}). It is the basis for Theorems~\ref{NLPCrel} and \ref{T1}.

\section{Cohomology}\label{CohoW}

In the previous section we found that for $l2$-MBQCs contextuality and computation hinge on the same algebraic structure. If we impose an ncHVM description on top of this structure, we obtain a contradiction; and if we do not impose it, we obtain a computation.  
This begs the question:  {\em{What precisely is this common algebraic structure underlying both parity-based contextuality proofs and measurement-based quantum computation?}} This is where cohomology comes in.\smallskip

Below, we build up the cohomological picture for deterministic, temporally flat MBQCs. The connection between MBQC and contextuality runs through state-dependent parity-based contextuality proofs. In Section~\ref{CohoCon}, we first introduce the cohomological description of  the state-independent counterpart. It is based on a chain complex ${\cal{C}}(E)$, and slightly simpler. We then progress to the state-dependent version, described by the relative chain complex ${\cal{C}}(E,E_0)$. In Section~\ref{CohoComp}, we explain the relation between cohomology in ${\cal{C}}(E,E_0)$ and MBQC output.

\subsection{Cohomology and contextuality}\label{CohoCon} 

We begin with the simpler state-independent parity proofs of contextuality, and then move on to their state-dependent cousins which are of more direct interest for MBQC. In all that follows we consider observables whose eigenvalues are all $\pm 1$. We denote these observables by $T_a$, a notation we now explain.

The basic object in the cohomological discussion of the parity proofs are chain complexes ${\cal{C}}(E)=(C_0,C_1,C_2,C_3)$ consisting of points (0-chains), edges (1-chains), faces (2-chains) and volumes (3-chains), and boundary maps $\partial$ between those chains. The observables $T_a$ forming the contextuality proof are associated with the edges $a\in E$ in the complex ${\cal{C}}(E)$. More precisely, each edge $a$ corresponds to an equivalence class $\{\pm T_a\}$ of observables, $a:=\{\pm T_a\}$. From each equivalence class $a$, one observable is picked and denoted as $T_a$. 

From the perspective of contextuality, the reason for considering the observables $T_a$ and $-T_a$ as equivalent is the following. If a parity-based contextuality proof can be based on some set of observables $\{T_a, a\in E\}$, then any signed set $\{(-1)^{\gamma(a)} T_a, \;\gamma(a) \in \mathbb{Z}_2, \forall a\in E\}$ produces an equivalent proof.  The signs $(-1)^{\gamma(a)}$ in the definition of the observables $T_a$ don't matter for the existence of contextuality proofs; and this leads us to consider the equivalence classes $\{\pm T_a\}$. We will return to this observation once we have set up the appropriate notation, right after Theorem~\ref{CPth1}.

The 1-chains $c_1\in C_1$ are linear combinations of the edges $a\in E$  with $\mathbb{Z}_2$ coefficients. 
The faces of ${\cal{C}}(E)$ are sets $f=(a_1,a_2,..,a_n)$ of edge labels $a_i$ of pairwise commuting operators $T_{a_i}$, such that for every face $f $ it holds that
\begin{equation}\label{ProdRel}
\prod_{a\in f} T_a = I (-1)^{\beta(f)},
\end{equation}
for a suitable function $\beta$ defined on the faces. We denote the set of faces by $F$, and the 2-chains $c_2 \in C_2$ are linear combinations of the faces $f\in F$ with coefficients in $\mathbb{Z}_2$. 

We can now define a boundary operator $\partial: C_2 \longrightarrow C_1$ via $\partial(f) = \sum_{a\in f} a$, for all $f\in F$, and extension from $F$ to $C_2$ by linearity. We can then also define a coboundary operator $d: C^1 \longrightarrow C^2$ in the usual way; i.e. for every 1-cochain $x \in C^1$ it holds that $d x(f):=x(\partial f)$, for all $f\in F$.\medskip

The function $\beta: C_2 \longrightarrow \mathbb{Z}_2$ plays a central role in the cohomological discussion of contextuality. Namely, assume that a non-contextual value assignment $\lambda$ exists, and as before write $\lambda(\cdot) = (-1)^{s(\cdot)}$. Then, Eq.~(\ref{ProdRel}) implies that $\beta(f) = \sum_{a\in f}s(a) = s(\partial f)$ for all $f\in F$. We may write this   in cochain notation as
\begin{equation}\label{betads}
\beta = ds.
\end{equation}
This equation may be interpreted as a constraint on the value assignment $s$, given $\beta$. But it may as well be regarded as a constraint on $\beta$. Namely, not all functions $\beta$ are of form Eq.~(\ref{betads}), for any 1-cochain $s$. Thus, a measurement setting based on ${\cal{C}}(E)$ is non-contextual only if $\beta = ds$ for some $s\in C^1$, or, equivalently, it is contextual if $\beta \neq ds$, for any $s\in C^1$.

\begin{figure}
\begin{center}
\begin{tabular}{lclcl}
(a) && (b) && (c)\\
\includegraphics[height=3.5cm]{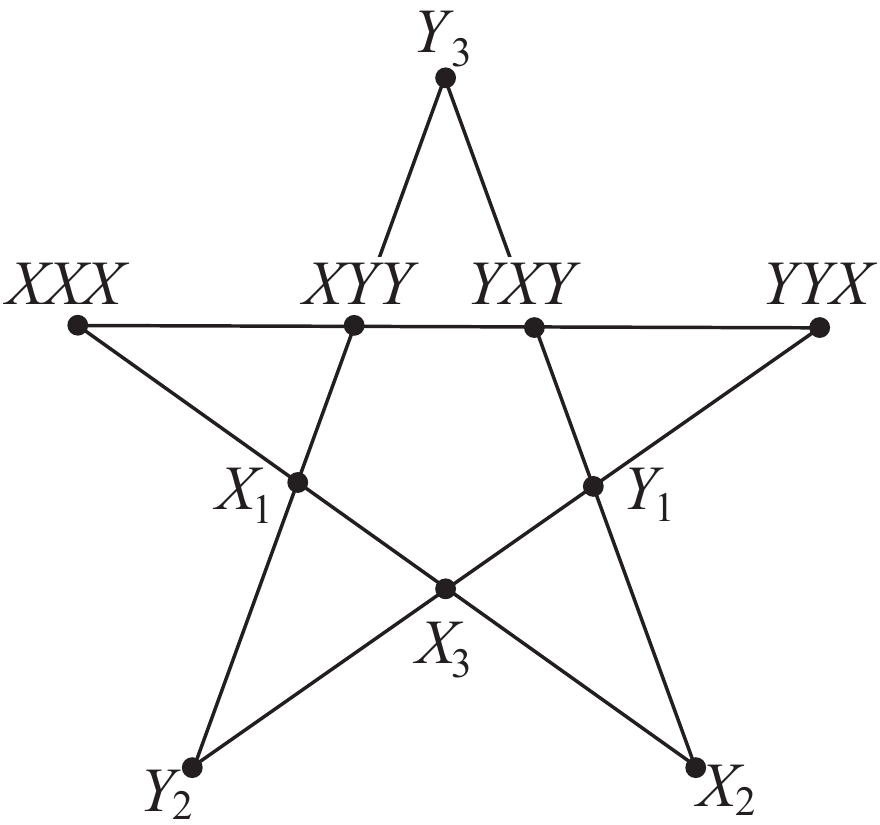} &&
\includegraphics[height=3.3cm]{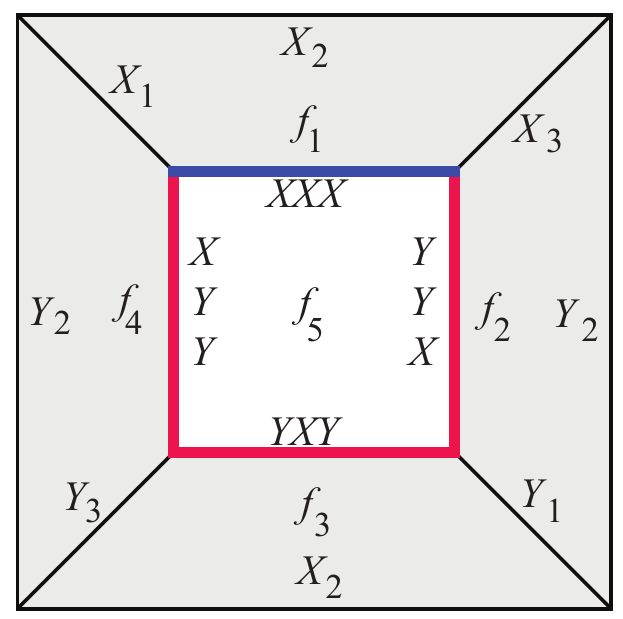}  &&
\includegraphics[height=3.3cm]{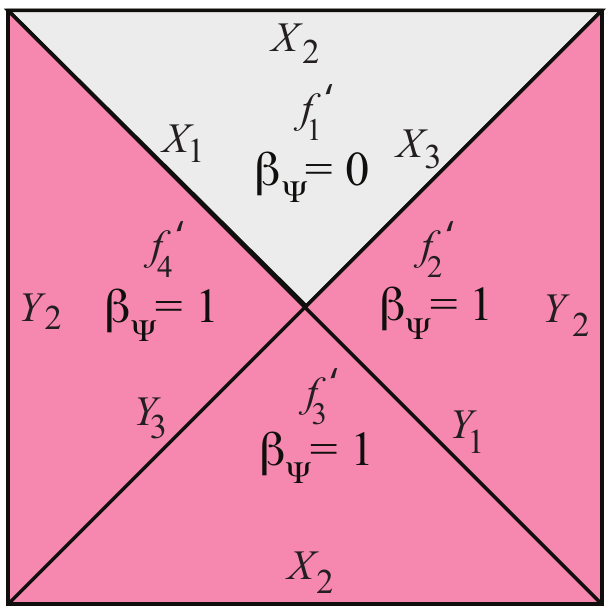} 
\end{tabular}
\caption{\label{MermSt} Mermin's star. (a) Standard representation. Each line represents a measurement context, composed of four commuting Pauli observables multiplying to $\pm I$. (b) Mermin's star re-arranged on a surface. The Pauli observables now correspond to edges, and each measurement context to the boundary of one of the four elementary faces. The exterior edges are pairwise identified. The colored edges carry a value assignment, resulting from the GHZ stabilizer. (c) Relative complex ${\cal{C}}(E,E_0)$. The edges corresponding to observables in the GHZ stabilizer are removed by contraction.}
\end{center}
\end{figure}

We will now slightly reformulate the last statement, to better bring out its cohomological nature. The function $\beta$ is by definition a 2-cochain. But in fact it is a 2-cocycle, $d\beta =0$ \cite{Coho}. Thus, we may express the above contextuality condition as follows.
\begin{Theorem}[\cite{Coho}]\label{CPth1}A set of measurements specified by the chain complex ${\cal{C}}(E)$ is contextual if for the cocycle class $[\beta] \in H^2({\cal{C}}(E),\mathbb{Z}_2)$ it holds that
$$[\beta] \neq 0.$$ 
\end{Theorem}
{\em{Remark:}} We observed above that no transformation $T_a \longrightarrow (-1)^{\gamma(a)} T_a$, $\forall a \in E$, affects the existence of contextuality proofs. We can now verify this statement in Theorem~\ref{CPth1}. At the level of the cocycle $\beta$, the transformations act as $\beta \longrightarrow \beta + d\gamma$. Hence, $[\beta] \longrightarrow [\beta]$. The parity proofs are thus indeed unchanged. We point out that the transformations discussed here---which we call gauge transformations---have a further use in characterizing MBQC output functions; see  Section~\ref{CohoComp}.

\medskip
Now let's consider the state-independent Mermin star in this framework.
The ten Pauli observables $T_a$ therein are assigned to the edges $a \in E$ in a chain complex ${\cal{C}}$; see Fig.~\ref{MermSt}b. For the five faces shown we have $\beta(f_1)=\beta(f_2) =\beta(f_3) = \beta(f_4)=0$, and $\beta(f_5)=1$. Further denote ${\cal{F}}:=\sum_{i=1}^5 f_i$, such that $\partial {\cal{F}} =0$ and $\beta({\cal{F}})=1$. Now assume Mermin's star were non-contextual. Then, $\beta=ds$ for some $s\in C^1$, and we have
$$
0 = s(0) = s(\partial {\cal{F}}) = ds({\cal{F}}) = \beta({\cal{F}}) = 1.
$$
Contradiction. Hence, Mermin's star is contextual.\medskip

We now seek a state-dependent version of Theorem~\ref{CPth1}, preferably formulated in an analogous way. This can be achieved by proceeding from the chain complex ${\cal{C}}(E)$ to a relative chain complex ${\cal{C}}(E,E_0)$. The quantum state $|\Psi\rangle$ now appears, and the set $E_0\subset E$ consists of those edges $a$ for which the corresponding operator $T_a$ has $|\Psi\rangle$ as an eigenstate, 
\begin{equation}\label{ES}
T_a|\Psi\rangle = (-1)^{\mu(a)}|\Psi\rangle,\;\text{with } \mu: E_0 \longrightarrow \mathbb{Z}_2.
\end{equation}
Geometrically, ${\cal{C}}(E,E_0)$ is obtained from ${\cal{C}}(E)$ by contracting the edges in $E_0$. Thereby, the faces of ${\cal{C}}(E)$ whose boundary lives entirely inside $E_0$ are  removed. Under this contraction, the boundary map $\partial$ changes to a relative boundary map $\partial_R$ defined by $\partial_R(f) = \sum_{a\in f\backslash E_0}a$.

Extending the above function $\mu$ to all of $E$ by setting $\mu(a):=0$ for all $a \not \in E_0$, we define a relative 2-cochain
\begin{equation}\label{betaPsi}
\beta_\Psi:= \beta + d\mu \mod 2.
\end{equation}
Again, $\beta_\Psi$ is a 2-cocycle. Also, $\beta_\Psi$ evaluates to zero on all faces with boundary entirely inside $E_0$, and it is thus a cocycle in the relative complex ${\cal{C}}(E,E_0)$.

Quantum mechanically, the measurement record in the context corresponding to any face $f\in F$ has to satisfy $s|_{f\cap E_0} = \mu |_{f \cap E_0}$, and $\beta(f) = s(\partial f)$. Then, from the above definitions it follows that
\begin{equation}\label{BetaPsi_s}
\beta_\Psi(f) = s(\partial_R f).
\end{equation}
Now assume a value assignment $s$ exists. It has to satisfy the condition Eq.~(\ref{BetaPsi_s}) for all faces $f\in F$ simultaneously. We may thus write the constraints on such a global value assignment $s$ as $ds=\beta_\Psi$, with $d$ now being the coboundary operator in the complex ${\cal{C}}(E,E_0)$. 

We thus have, in complete analogy with the state-independent case, the following result.
\begin{Theorem}[\cite{Coho}]\label{CPth2}A set of measurements and a quantum state $|\Psi\rangle$ specified by the chain complex ${\cal{C}}(E,E_0)$ are contextual if for the cocycle class $[\beta_\Psi] \in H^2({\cal{C}}(E,E_0),\mathbb{Z}_2)$ it holds that
$$[\beta_\Psi] \neq 0.$$ 
\end{Theorem}
{\em{Example, Part II.}} We now apply this to the example of the state-dependent Mermin star. Four faces remain in ${\cal{C}}(E,E_0)$ after contraction of $E_0$ in ${\cal{C}}(E)$, $f_1',.., f_4'$. We have
 $\beta_\Psi(f_1')=0$, $\beta_\Psi(f_2')= \beta_\Psi(f_3')=\beta_\Psi(f_4')=1$.
Denote ${\cal{F}}'=\sum_{i=1}^4 f_i'$ such that the relative boundary of ${\cal{F}}'$ vanishes, $\partial_R {\cal{F}}'=0$, and $\beta_\Psi({\cal{F}}')=1$. 

Now assume that the state-dependent Mermin star is non-contextual. Then, $\beta_\Psi=ds$ for some 1-cochain  $s\in C^1({\cal{C}}(E,E_0),\mathbb{Z}_2)$. And thus 
\begin{equation}\label{cohoCount}
1= \beta_\Psi({\cal{F}}') = s({\partial {\cal{F}}'}) = s(0) =  0.
\end{equation}
Contradiction. Hence the state-dependent Mermin star is contextual.

Eq.~(\ref{cohoCount}) is the cohomological version of Eq.~(\ref{sdMS}). It describes the exact same system of linear constraints.

\subsection{Cohomology and computation}\label{CohoComp} 

Recall from Section~\ref{MBQCrev} that in MBQC there are two measurable observables at each physical site $i$, $O_i[q_i]$, $q_i \in \mathbb{Z}_2$.  To make use of the cohomological formalism, we now denote these observables as
\begin{equation}\label{OT}
O_i[0]=T_{a_i},\; O_i[1]=T_{\overline{a}_i},\;\; \forall i=1,..,n.
\end{equation}
We define the notion of an input group to import the classical processing relation Eq.~(\ref{CCR_in}) into our cohomological picture. The input group is 
$Q =\langle \textbf{i}_j,\;j=1,..,m\rangle \cong \mathbb{Z}_2^m$. The generators of $Q$ act on the observables of Eq.~(\ref{OT}) as 
\begin{equation}\label{cflip}
\begin{array}{rrl}
\textbf{i}_j(a_i)=(a_i),&\textbf{i}_j(\overline{a}_i)=(\overline{a}_i),& \text{if $S_{ij}=0$},\\
\textbf{i}_j(a_i)=(\overline{a}_i),&\textbf{i}_j(\overline{a}_i)=(a_i),& \text{if $S_{ij}=1$}.
\end{array}
\end{equation}
Denoting by ${\cal{E}}_\text{e}$ a reference context corresponding to the trivial input $\text{e} \in Q$, ${\cal{E}}_\text{e} := \{a_j,\,j = 1,..,n\}$, and by ${\cal{E}}_\textbf{i}$ the measurement context for any input $\textbf{i} \in Q$, then, with the definitions Eq.~(\ref{OT}) and (\ref{cflip}), the relation
\begin{equation}\label{Iact}
{\cal{E}}_\textbf{i} = \textbf{i}({\cal{E}}_\text{e}):=\{\textbf{i}(a_j),\,j=1,.., n\}
\end{equation}
reproduces the classical side processing relation Eq.~(\ref{CCR_in}) in the limit of temporally flat MBQCs, $T=0$. This is the limit we are presently interested in.

We have thus far represented computational input by a group $Q$ that maps the complex ${\cal{C}}(E,E_0)$ to itself, and we now turn to the computational output. In terms of the above sets ${\cal{E}}_\textbf{i}$, the classical side processing relations for output, Eq.~(\ref{CCR_out}), read
\begin{equation}\label{CCR_out_2}
o(\textbf{i}) = \sum_{a\in {\cal{E}}_\textbf{i}} s(a) \mod 2,\;\; \forall \textbf{i} \in Q.
\end{equation}
We note that for any $\textbf{i} \in Q$, the observables $T_a$, $a\in {\cal{E}}_\textbf{i}$, pairwise commute.  Furthermore, in the setting of deterministic computation, the input group $Q$ (equivalently, the matrix $S$ in Eq.~(\ref{CCR_in})) is chosen such that the MBQC resource state $|\Psi\rangle$ is an eigenstate of all observables $\prod_{a \in {\cal{E}}_\textbf{i}} T_a $. That is, $\prod_{a \in {\cal{E}}_\textbf{i}} T_a =\pm T_x$, with $x\in E_0$; cf. Eq.~(\ref{ES}). Therefore, the edges $a\in {\cal{E}}_\textbf{i}$ form the boundary of a face $f_\textbf{i}$ in the contracted complex ${\cal{C}}(E,E_0)$, i.e., $f_\textbf{i} \in C_2({\cal{C}}(E,E_0))$ satisfies ${\cal{E}}_\textbf{i} = \{\partial_R f_\textbf{i}\}$. Finally, with Eq.~(\ref{Iact}), ${\cal{E}}_\textbf{i} = \{\textbf{i} (\partial_R f_\text{e})\}$, and the face $f_\text{e}$ corresponds to ${\cal{E}}_\text{e}$. Therefore, Eq.~(\ref{CCR_out_2}) can be rewritten in cohomological notation as
$$
o(\textbf{i}) = s(\textbf{i}(\partial_R f_\text{e})),
$$
where $s$ is the measurement record for the observables in ${\cal{E}}_\textbf{i}$. 

Inserting Eq.~(\ref{BetaPsi_s}) into the last equation, we obtain the following result.
\begin{Theorem}[\cite{CohoMBQC}]\label{ObeT}
The function $o: Q \longrightarrow \mathbb{Z}_2$ computed in a given deterministic and temporally flat $l2$-MBQC is related to the cocycle $\beta_\Psi \in C^2({\cal{C}}(E,E_0))$ via
\begin{equation}\label{obeta}
o(\textbf{i}) = \beta_\Psi(\textbf{i}(f_\text{e})),\; \forall \textbf{i} \in Q.
\end{equation}
\end{Theorem}
This relation between the computational output $o$ and the 2-cocycle $\beta_\Psi$ is the main result of this section. It has been established in greater generality in \cite{CohoMBQC} (Theorem~4 therein), but we don't need the additional generality here. Theorem~\ref{ObeT} is the computational counterpart to Theorem~\ref{CPth2} in Section~\ref{CohoCon}. Both results together establish that a single cohomological object, the cocycle $\beta_\Psi$, governs contextuality and computational output in MBQC. Jointly, Theorems~\ref{CPth2} and \ref{ObeT} thus flesh out the Diagram~(\ref{Triangle}).\medskip

\begin{figure}
\begin{center}
\includegraphics[width=8cm]{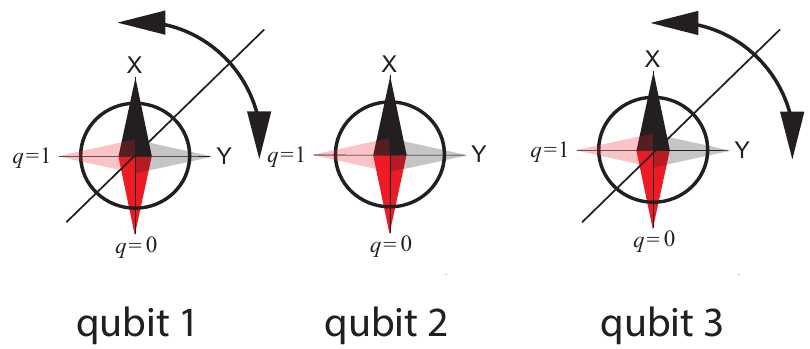}
\caption{\label{GHZsymm}Input group of the GHZ-MBQC. Displayed is the action of the element $\textbf{i}_1$ of the input group $Q = \mathbb{Z}_2\times \mathbb{Z}_2$. As described by Eq.~(\ref{Qghz}), for qubits 1 and 3, $X$ and $Y$ are interchanged under the given input, and the reference context $(X_1,X_2,X_3)$ is thereby changed into $(Y_1,X_2,Y_3)$. }
\end{center}
\end{figure}

{\em{Example, Part III.}} For the GHZ-MBQC, Eq.~(\ref{obeta}) may be explicitely verified by inspecting Fig.~\ref{MermSt}c. 
The reference context is ${\cal{E}}_{\text{e}}=(a_{X_1},a_{X_2},a_{X_3})$, hence $f_\text{e} = f'_1$, w.r.t. the labeling of Fig.~\ref{MermSt}c. The input group is $Q=\mathbb{Z}_2\times \mathbb{Z}_2$. Its two generators $\textbf{i}_1, \textbf{i}_2$ are related to the input bits $y$, $z$ of the OR-gate via $y \mapsto \textbf{i}_1,\; z \mapsto \textbf{i}_2$, and Eq.~(\ref{cflip}) becomes
\begin{equation}\label{Qghz}
\begin{array}{rl}
\textbf{i}_1: &a_{X_1} \leftrightarrow a_{Y_1}, \;  a_{X_3} \leftrightarrow a_{Y_3}, \; a_{X_2}  \circlearrowright,\; a_{Y_2} \circlearrowright,\\
\textbf{i}_2: &a_{X_2} \leftrightarrow a_{Y_2}, \;  a_{X_3} \leftrightarrow a_{Y_3}, \; a_{X_1}  \circlearrowright,\; a_{Y_1} \circlearrowright.
\end{array}
\end{equation}
We may now verify in the cohomological calculus established above that this action does indeed lead to the execution of the OR-gate in the corresponding GHZ-MBQC. For example, if $y=z=0$ then $\textbf{i}=\text{e}$, and $\textbf{i}(f_1') =f_1'$; and thus $o(0,0) = \beta_\Psi(f_1') = 0 = \text{OR}(0,0)$. Further, if $y=1$ and $z=0$, then $\textbf{i} =\textbf{i}_1$, and $\textbf{i}_1(f'_1) = f_3'$. Thus, $o(1,0) = \beta_\Psi(f_3')=1=\text{OR}(1,0)$. The other two cases are analogous. See Fig.~\ref{GHZsymm} for illustration of the action of the input group given by Eq.~(\ref{cflip}).\medskip

One point remains to be discussed. When comparing Theorems~\ref{CPth2} and \ref{ObeT}, we notice a difference. Theorem~\ref{CPth2} invokes the cohomology class $[\beta_\Psi]$ whereas Theorem \ref{ObeT} invokes the cocycle $\beta_\Psi$ itself. Only the former theorem is therefore truly topological. This prompts the question: {\em{Is there an operationally meaningful way of grouping the MBQC output functions $o$  into equivalence classes $[o]$ that depend only on $[\beta_\Psi]$?}}

That is indeed the case. The equivalence classes $[o]$ of MBQC output functions are motivated and constructed as follows. We note that the signs in the observables $\{T_a,\,a\in E\backslash E_0\}$ are a mere convention. If an observable $T_a$, for some $a\in E\backslash E_0$, is measured in a given MBQC, then a measurement of $-T_a$ is exactly as hard, because the corresponding projectors are the same. To obtain one measurement from the other, only the labels of the two pointer positions of the measurement device need to be switched. Therefore, the change
\begin{equation}\label{GT}
T_a \longrightarrow (-1)^{\gamma(a)} T_a, \;\forall a\in E\backslash E_0,
\end{equation}
for any cochain $\gamma: C_1(E,E_0) \longrightarrow \mathbb{Z}_2$ is an equivalence transformation, or, as it is also called, a gauge transformation. 

Yet, these transformations have an effect. The cocycle $\beta_\Psi$ changes, namely
$$
\gamma: \beta_\Psi \mapsto \beta_\Psi + d\gamma.
$$
And thus, by Theorem~\ref{ObeT}, the outputted function $o$ changes too. Functions obtained from one another through such a transformation should be considered computationally equivalent, as was argued above. It is thus meaningful to group MBQC output functions $o$ into equivalence classes
$$
[o(\cdot)]:=\{(\beta_\Psi+d\gamma)(\cdot  f_\text{e}),\;\forall \gamma \in C^1({\cal{C}}(E,E_0))\}.
$$
With this definition, Theorem~\ref{ObeT} has the following corollary.
\begin{Cor}\label{CohoCompCor}
For each deterministic and temporally flat $l2$-MBQC, the equivalence class $[o]$ of output functions is fully determined by $[\beta_\Psi]$.
\end{Cor}
Thus, the gauge-invariant information in an MBQC output function is contained in the same cohomological information that also provides the contextuality proof. 
\medskip

{\em{Example, Part IV.}} In the GHZ-MBQC, we may flip  $Y_3 \longrightarrow - Y_3$. In result, the new computed function is an AND. Therefore, AND and OR are equivalent wrt. MBQC. Considering the whole set of equivalence transformations for this example, we find that there are two equivalence classes of functions on two bits, the non-linear Boolean functions and the linear ones. Each member of the former class  boosts the classical control computer of MBQC to computational universality, whereas the second class has no effect on the computational power at all. 

From the cohomological perspective, $H^2({\cal{C}}(E,E_0),\mathbb{Z}_2) = \mathbb{Z}_2$, i.e. there are two equivalence classes of cocycles $\beta_\Psi$. The trivial class corresponds to the linear Boolean functions on two bits and the non-trivial class to the non-linear Boolean functions.

\subsection{On the probabilistic case}

In the previous sections we focussed to deterministic MBQC. Indeed, powerful deterministic quantum algorithms do exist, notably for the Discrete Log problem \cite{MZ}. However, most known quantum algorithms are probabilistic, i.e., they succeed with a probability smaller than one.   A cohomological treatment of probabilistic MBQCs is given in \cite{CohoMBQC}, based on group cohomology. Here we are content with alerting the reader to the additional layer of difficulty posed by the probabilisitic case.\smallskip

Let's trace the restriction to deterministic MBQCs back to its origin. In Theorem~\ref{ObeT}, the central result on the computational side, it is present through the cocycle $\beta_\Psi \in C^2({\cal{C}}(E,E_0),\mathbb{Z}_2)$. This cocycle is defined in Eq.~(\ref{betaPsi}), in terms of the cocycle $\beta \in C^2({\cal{C}}(E),\mathbb{Z}_2)$ and the value assignment $\mu: E_0 \longrightarrow \mathbb{Z}_2$. The value assignment $\mu$ in turn refers to eigenvalues of certain observables related to computational output, of which the resource state $|\Psi\rangle$ is an eigenstate; cf. Eq.~(\ref{ES}). 

In the probabilistic case, the value assignment $\mu$ does in general not exist. Hence, $\beta_\Psi$ is not defined, and we cannot have straightforward probabilistic counterparts of Theorems \ref{CPth2} and \ref{ObeT}.\smallskip

But the problem is not merely technical; it is conceptual. Consider our running example of the GHZ-MBQC, which executes an OR-gate with certainty. As soon as probabilistic computations are admitted, we may as well say that it evaluates the constant function $y\equiv 1$ with an average success probability of 75 percent. In fact, the same computation executes any 2-bit Boolean function, except $\neg \text{OR}$, with some nonzero probability of success. How can we then say that one particular function is computed while all others are not?

 Key to the solution is a group $G$ of symmetry transformations that extends the input group $Q$, in the group-theoretic sense. $G$ maps the complex ${\cal{C}}(E,E_0)$ to itself, acting on the observables $T_a$, $a\in E\backslash E_0$ via
\begin{equation}\label{PhiDef}
g(T_a) = (-1)^{\tilde{\Phi}_g(a)}T_{ga},\;\;\forall g\in G.
\end{equation}
Therein, the phase function $\tilde{\Phi}$ is, per construction, a 1-cocycle in group cohomology.

There is a further condition on $G$. Namely, the action Eq.~(\ref{PhiDef}) of $G$ on the set of observables $\{\pm T_a, a\in E\backslash E_0\}$ induces an action on the output function $o$, and we require $o$ to be invariant under this action. It turns out that, given $G$, this invariance condition constrains $o$ up to an additive constant \cite{CohoMBQC}. Thus, the output function $o$ is {\em{defined}} through a symmetry group.

Furthermore, $o$ can be expressed in terms of the phase function $\tilde{\Phi}$, and a contextuality proof can be given in terms of a group cohomology class derived from $\tilde{\Phi}$. In result, Theorems~\ref{CPth2} and \ref{ObeT} have counterparts in the probabilistic case. They are given as Theorem 5 in \cite{Coho} and Theorem 6 in \cite{CohoMBQC}, respectively. The probabilistic counterpart of Corollary~\ref{CohoCompCor} is Corollary~2 in \cite{CohoMBQC}.

\section{Temporal order}\label{TO}

The connection between contextuality and $l2$-MBQC described by Theorem~\ref{NLPCrel} is completely general. It applies to deterministic and probabilistic measurement-based computations, as well as temporally flat and temporally ordered ones. It is only the cohomological description of this connection that is presently restricted to temporally flat computations. This is a technical limitation, and the purpose of this section is to outline an approach for overcoming it.  

The idea is to not change the cohomological description at all, but to enlarge the complex ${\cal{C}}(E,E_0)$ by additional observables which take care of the temporal ordering. We illustrate this approach with the setting of the ``iffy'' proof \cite{Exa}.

In Section~\ref{Iffy1} we review the iffy contextuality proof, largely following the original exposition \cite{Exa}. We then explain why the signature feature of iffiness is incompatible with applications to MBQC. In Section~\ref{Iffy2} we present a cohomological contextuality proof for the iffy scenario that is MBQC-compatible. This proof includes temporal order, yet is covered by Theorem~\ref{CPth2} without any modification.

\subsection{The ``iffy" contextuality proof}\label{Iffy1}

To get started, we require a simple example for a contextuality proof with temporal order, a counterpart to the non-adaptive GHZ proof. Luckily, Ref.~\cite{Exa}, Section 6, offers one; in fact, it offers a whole family of examples. We begin by writing them in a stabilizer notation that suits our purpose. 

The examples consist of a three-qubit resource state $|\Psi\rangle$, and local measurement settings for the three qubits. For any even integer $N$, choose
$$
|\Psi\rangle \sim |00\rangle |\nu\rangle + | 11\rangle |\omega\rangle,
$$
where
$$
\begin{array}{rcl}
|\nu\rangle &=& \cos \frac{\lambda}{2}|0\rangle + \sin\frac{\lambda}{2}|1\rangle,\\
|\omega\rangle &=&\sin \frac{\lambda}{2}|0\rangle + \cos\frac{\lambda}{2}|1\rangle,
\end{array}
$$
and $\lambda = \pi/2 - \pi/N$. This defines the resource state. Now the measurements: qubit 3 will be measured in the eigenbasis of $X$ or $Y$, and qubits 1 and 2 will be measured in the eigenbases of any of the operators
\begin{equation}\label{XkDef}
X_k:= \cos\left( k\frac{\pi}{N}\right) X + \sin\left( k\frac{\pi}{N}\right)Y,\;\; \forall k =0 ... 2N-1.
\end{equation}
Note that $X_{N+k} = - X_k$, such that we really only need the observables $X_0$, .. , $X_{N-1}$.

Denote by $P_{y,\pm}$ the projector on the eigenstate of $Y$ with positive and negative eigenvalue, respectively, and define the operators
\begin{equation}\label{tauX}
\begin{array}{rcll}
\tau_k &: =& X_{N-1-k}^{(1)} \otimes X_{k}^{(2)}\otimes P_{y,+}^{(3)} +  X_{N+1-k}^{(1)} \otimes X_{k}^{(2)}\otimes P_{y,-}^{(3)},& k=0,.., N-1,\\
\overline{X}_k &:=& X^{(1)}_{N-k}\otimes X^{(2)}_k \otimes X^{(3)},& k=0,.., N-1.
\end{array}
\end{equation}
By direct calculation, we can verify that
\begin{subequations}\label{Stab}
\begin{align}\label{StabX}
\overline{X}_k |\Psi\rangle &=- |\Psi\rangle,\; \forall k,\\ 
\label{StabY}
\tau_k |\Psi\rangle &= - |\Psi\rangle,\; \forall k.
\end{align}
\end{subequations}
The measurement strategies considered in the contextuality proof have temporal order. Namely, first qubit 3 is measured, in the $X$ or $Y$ basis. In the latter case, the further choice of the measurement bases for qubits 1 and 2 depends on the outcome of the measurement at 3.

From Eq.~(\ref{Stab}) we can read off  the constraints on the non-contextual hidden variable model, which are provided in \cite{Exa}. Denote by $a_k$ and $b_k$ the binary measurement outcomes on qubits 1 and 2, respectively, given the measured observable $X_k$, and by $c_0$ ($c_1$) the outcome on qubit 3 if the measured observable is $X$ ($Y$). If these values form the value assignment of an ncHVM, they must satisfy the constraints
\begin{equation}\label{ValAss}
\begin{array}{rcll}
a_i \oplus b_j \oplus c_0 &=& 0, & \forall i,j\;\text{s.th. } i+j =0,\\
a_i \oplus b_j \oplus c_0 &=& 1, & \forall i,j\;\text{s.th. } i+j =N,\\ \\
a_i \oplus b_j  &=& 0, & \forall i,j\;\text{s.th. } i+j +(-1)^{c_1} =0,\\
a_i \oplus b_j  &=& 1, & \forall i,j\;\text{s.th. } i+j +(-1)^{c_1} =N.
\end{array}
\end{equation}
The contextuality proof proceeds from there, as usual, by adding up equations mod 2. This will be discussed below.\medskip

We now show how Eq.~(\ref{ValAss}) are derived from the stabilizer relations Eq.~(\ref{Stab})\footnote{The original derivation of Eq.~(\ref{ValAss}) in \cite{Exa} uses a different formalism which we do not reproduce here.}. The two relations at the top of Eq.~(\ref{ValAss}) follow straightforwardly from Eq.~(\ref{StabX}); here we focus on the relations at the bottom of Eq.~(\ref{ValAss}), which derive from Eq.~(\ref{StabY}).

First, for the observables of Eq.~(\ref{tauX}), with Eq.~(\ref{Stab}) we have the following values
\begin{equation}
s_{\tau_k}= s_{\overline{X}_k} =1,\;\;k=0,.., N-1.
\end{equation}
corresponding to eigenvalues $(-1)^1=-1$.  Now consider separately the two cases of $c_1=0$ and $c_1=1$, respectively.

Case I:  $c_1=0$. We now want to argue that, in this case, the observables
$
\tau_k(0) =  X_{N-1-k}^{(1)} \otimes X_{k}^{(2)}
$
are also assigned the value 1, 
$$
s_{\tau_k(0)} =1, \;\; k=0,..,N-1.
$$
The argument is as follows. If $c_1=0$, then this fact could be established by measuring $Y^{(3)}$. According to quantum mechanics, the post-measurement state would be $|y,+\rangle:=P_{y,+}^{(3)}|\Psi\rangle$. For this state it holds that 
\begin{equation}\label{eqChain}
\tau_k(0)|y,+\rangle = \tau_k|y,+\rangle =  \tau_k P_{y,+}|\Psi\rangle =   P_{y,+} \tau_k|\Psi\rangle = - P_{y,+} |\Psi\rangle =-|y,+\rangle. 
\end{equation}
For later reference, note that in the above chain of equalities we have used the properties
\begin{equation}\label{ref}
 \tau_k(0)P_{y,+} = \tau_k P_{y,+}\mbox{ and } [\tau_k,P_{y,+}]=0.
\end{equation}
By Eq.~(\ref{eqChain}), $s_{\tau_k(0)} =1$, for all $k$, as claimed. Further, by standard arguments, $s_{\tau_k(0)} =a_{N-1-k} \oplus b_k$. Combining the last two statements,
$$
a_{N-1-k} \oplus b_k = 1,\;\;\forall k=0,.., N-1.
$$
This provides the lower part of Eq.~(\ref{ValAss}) for the case of $c_1=0$.

Case II:  $c_1=1$. A completely analogous argument establishes the bottom half of Eq.~(\ref{ValAss}) for $c_1=1$.\smallskip

Eq.~(\ref{ValAss}) is thus established as a set of constraints that any value assignment $\{a_k,b_l,c_m\}$ needs to satisfy. We now complete the proof, focussing on Case I, $c_1=0$. Case II is analogous. 

We assume that a value assignment exists. From the upper half of Eq.~(\ref{ValAss}), we pick the equation $a_0 +b_0+c_0 \mod 2=0$, and the equations $a_k+b_{N-k}+c_0 \mod 2 =1$, for $k=1,..,N-1$. From the lower half we pick the equations $a_l + b_{N-1-l} =1$, for $l=0,..,N-1$. Summing those equations, we obtain $N c_0 +2\sum_{k=0}^{N-1}(a_k+b_k) = 2N-1\;\;(\text{mod}\; 2)$. Since $N$ is even, this is a contradiction. $\Box$ \medskip

Now that we have presented the iffy contextuality proof, let's take a step back and ask two questions.

(1) {\em{Where is temporal order in this contextuality proof?}}---Suppose one wants to test the correlations of Eq.~(\ref{tauX}) through local measurement. The correlations are labeled by an integer $k \in \mathbb{Z}_N$, and a further binary integer $l\in \mathbb{Z}_2$ that decides whether qubit \#3 is measured in the $X$-basis ($l=0$) or in the $Y$-basis $(l=1)$. Given the input $(k,l)$, the pattern of local measurements to test the correlations is fully specified. Therein, if $l=1$, the measurement basis for qubit \#1 depends on the outcome $c_1$ obtained on qubit \#3, cf. Eq.~(\ref{tauX}), upper line. Thus, qubit \#1 must be measured {\em{after}} qubit \#3. This is the same temporal ordering due to adaptive measurement as occurs in MBQC.\smallskip

\begin{figure}
\begin{center}
\begin{tabular}{lcl}
(a) && (b)\\
\includegraphics[width=4cm]{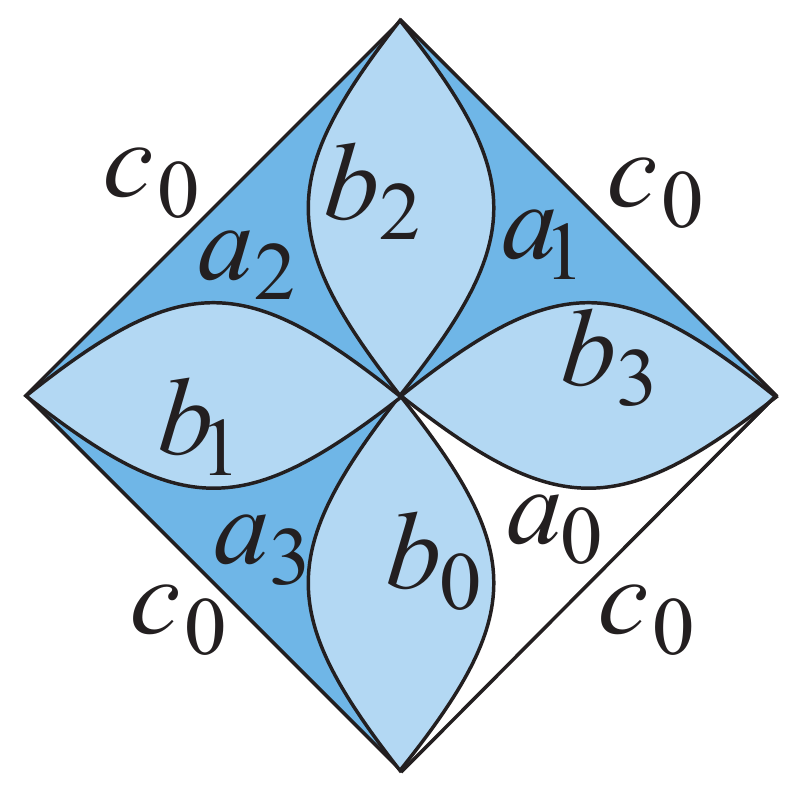} & &\includegraphics[width=4cm]{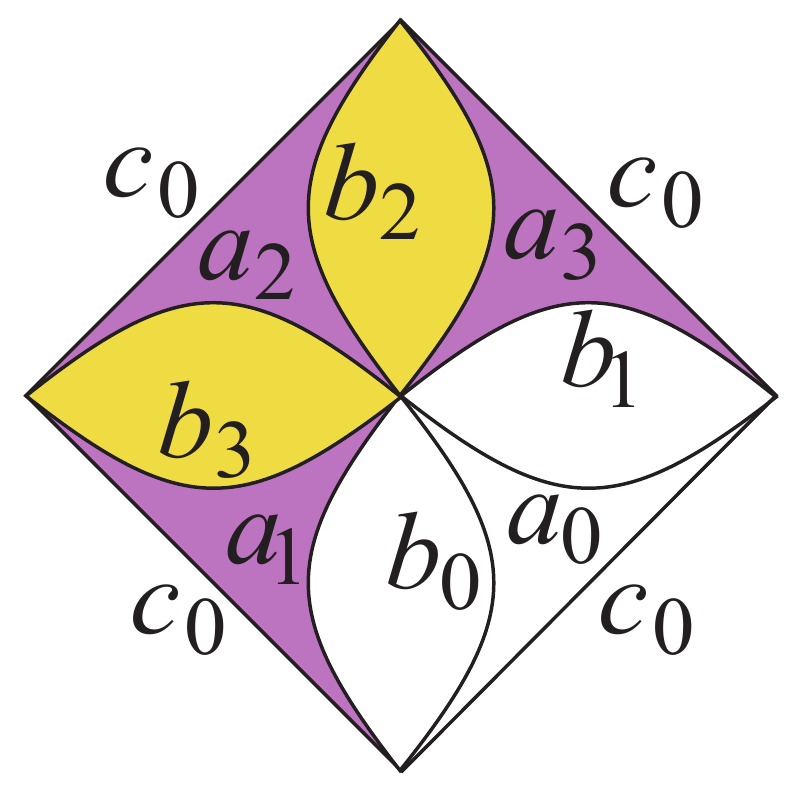}
\end{tabular}
\caption{\label{topol}Chain complexes in the iffy proof, for $N=4$. (a) the complex ${\cal{C}}^{(0)}$ for $c_1=0$ and (b) the complex ${\cal{C}}^{(1)}$ for $c_1=1$. In either case, the four edges labeled ``$c_0$'' correspond to the same observable $X^{(3)}$, and are identified.  The faces $f$ on which $\beta_\Psi(f)=1\, (0)$ are shown in color (white).}
\end{center}
\end{figure}

(2) {\em{Is the iffy proof topological?}}---Yes, but with a caveat. The value assignment for $c_1$ is not part of the topological description. Instead there are two separate topological descriptions, one for $c_1=0$ and one for $c_1=1$. They are depicted in Fig.~\ref{topol}, (a) the complex ${\cal{C}}^{(0)}$ for $c_1=0$ and (b) the complex ${\cal{C}}^{(1)}$ for $c_1=1$. In both cases there is a surface ${\cal{F}}^{(c_1)}$ comprising all of the faces displayed. Those surfaces have the property that $\partial {\cal{F}}^{(c_1)} =0$. In both cases it holds that $\beta_\Psi^{(c_1)}({\cal{F}}^{(c_1)})=1$, which, together with the former statement, implies that $\left[\beta_\Psi^{(c_1)}\right]\neq 0$, $\forall c_1\in \mathbb{Z}_2$. The iffy proof thus has two cohomological parts, conditioned by the value of $c_1$,
\begin{equation}\label{IP}
\text{Iffy\,Proof} =\left\{ \mathbb{Z}_2 \ni c_1 \mapsto \left({\cal{C}}^{(c_1)}, \beta_\Psi^{(c_1)}\right)\right\}.
\end{equation}

The conditioning on $c_1$ is in the way of using the iffy proof as a template for describing temporally ordered MBQCs. To see why this is so, let's recap the earlier topological proofs. There, the assumption of a noncontextual  value assignment $s$ is contradicted by $[\beta_\Psi]\neq 0$, and $\beta_\Psi$ is an object that is well-defined in quantum mechanics. Beyond the contextuality witness (see Theorem~\ref{CPth2}), $\beta_\Psi$ also contains the function computed in MBQC (see Theorem~\ref{ObeT}).

The counterpart of $\beta_\Psi$ in the present iffy proof is the quantum-classical hybrid structure given by Eq.~(\ref{IP}). It consists of the quantum-mechanically valid parts ${\cal{C}}^{(c_1)}$, $\beta_\Psi^{(c_1)}$, and one element, $c_1$, of the non-contextual value assignment, so far assumed to exist. (Recall that ruling out the existence of such a value assignment is the very purpose of the contextuality proof.) Unlike $\beta_\Psi$ in the former cases, as a whole this hybrid object is not compatible with quantum mechanics. It is thus not suitable to base a description of MBQC on. Now that we have understood this, we seek to modify the iffy proof such that it becomes compatible with measurement-based quantum computation.

\subsection{Deiffifying the iffy proof}\label{Iffy2}

Here we present a topological contextuality proof for the above iffy scenario that uses a complex of the type defined in \cite{Coho}. The proof works in completely the same way as in the temporally flat scenarios it was previously applied to.

\begin{figure}
\begin{center}
\includegraphics[width=8cm]{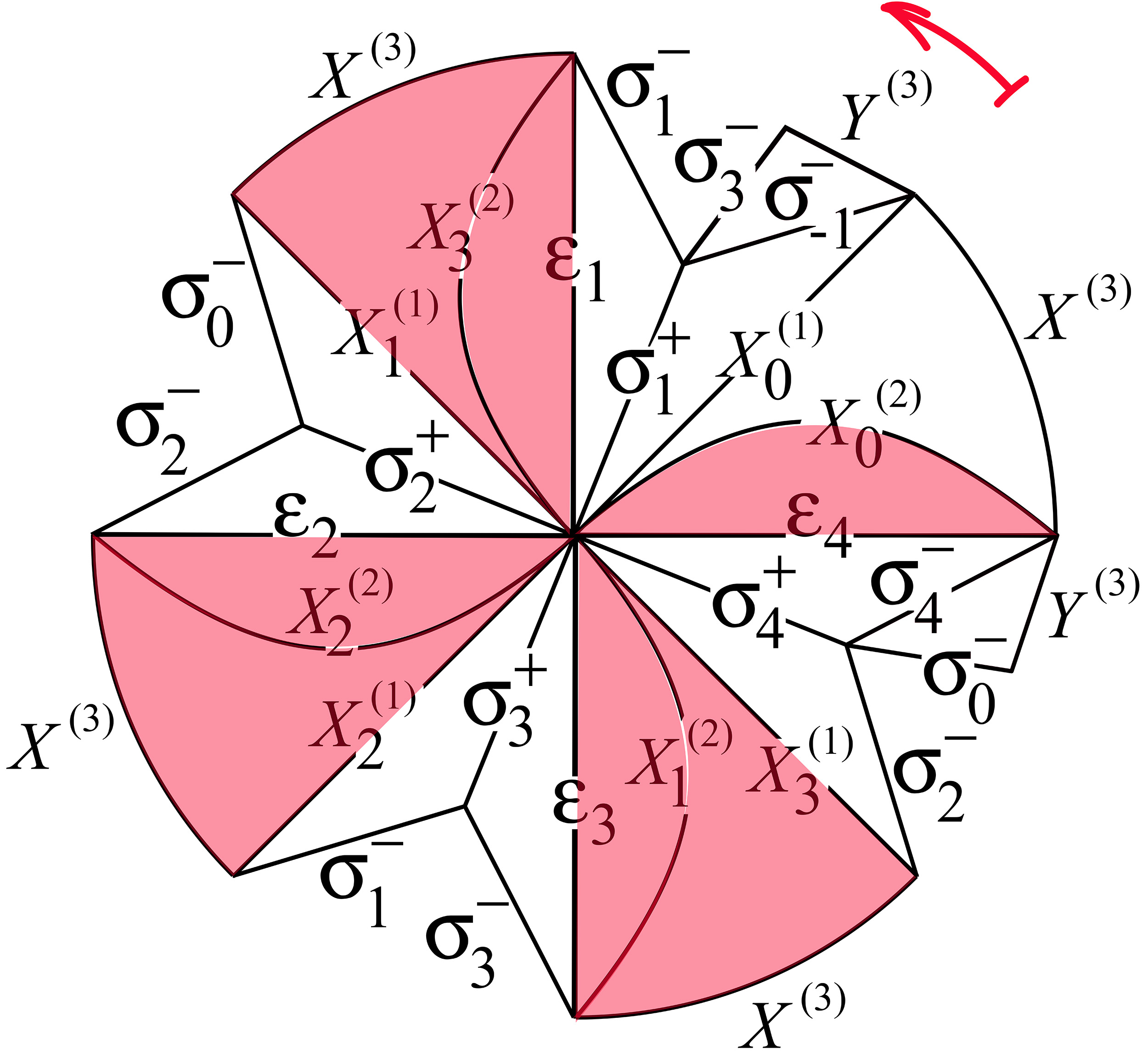}
\caption{\label{Comp1}Complex for the cohomological contextuality proof of the iffy scenario. There are four edges corresponding to $X^{(3)}$, and two each for $\sigma^\pm_k$, for various values of $k$, and for $Y^{(3)}$. Such edges are identified. The faces $f$ coloured in red have $\beta_\Psi(f)=1$, and the white faces $g$ have $\beta_\Psi(g)=0$.}
\end{center}
\end{figure}

We define a couple of extra observables, for all $k\in \mathbb{Z}_{2N}$,
\begin{subequations}\label{EpSig}
\begin{align}
\epsilon_k &:= \frac{I^{(3)}+Y^{(3)}}{2} \otimes X^{(1)}_{k-1} + \frac{I^{(3)}-Y^{(3)}}{2} \otimes X^{(1)}_{k+1},\\
\sigma^+_k &:= \frac{I^{(3)}+Y^{(3)}}{2} \otimes X^{(1)}_{k-1} + \frac{I^{(3)}-Y^{(3)}}{2} \otimes I^{(1)},\\
\sigma^-_k &:= \frac{I^{(3)}+Y^{(3)}}{2} \otimes I^{(1)} + \frac{I^{(3)}-Y^{(3)}}{2} \otimes X^{(1)}_{k+1}.
\end{align}
\end{subequations}
These are correlated observables on qubits \#1 and \#3. They can also be considered as unitary gates in which qubit \#3 is the control and qubit \#1 the target.  This is how the original iffiness enters into our topological proof, but in a fully quantum fashion. 

The stabilizer relations Eq.~(\ref{Stab}) can be expressed in terms of the observables defined in Eq.~(\ref{EpSig}). (only the first relation changes),
\begin{subequations}\label{StabRel2}
\begin{align}
\label{SR2a}
\epsilon_{N-k} \otimes X^{(2)}_k \, |\Psi\rangle &= - |\Psi\rangle,\\
\label{SR2b}
X_{N-k}^{(1)} \otimes X^{(2)}_k X^{(3)}\, |\Psi\rangle &=- |\Psi\rangle. 
\end{align}
\end{subequations}
Further, the observables $\epsilon_k$, $\sigma^\pm_k$ satisfy the following {\em{recoupling relations}}:
\begin{subequations}\label{esRel}
\begin{align}
\label{RelA}
\epsilon_k &= \sigma_k^+ \sigma_k^-,\\
\label{RelB}
X^{(1)}_k &= \sigma^+_{k+1}\sigma^-_{k-1},\\
\label{RelC}
-Y^{(3)} &= \sigma^+_k \sigma^+_{N+k},\\
\label{RelD}
Y^{(3)} &= \sigma^-_k \sigma^-_{N+k}.
\end{align}
\end{subequations}
Finally, we note the commutation relations
\begin{subequations}\label{CommRel}
\begin{align}
[\sigma^+_k,\sigma^-_l]&=0,\;\; \forall k,l \in\mathbb{Z}_{2N},\\
[\sigma^\pm_k,Y^{(3)}]&=0,\;\; \forall k, \in\mathbb{Z}_{2N}.
\end{align}
\end{subequations}
With these relations, the complex shown in Fig.~(\ref{Comp1}) is well-composed. I.e., all faces correspond to triples of commuting operators that multiply to $\pm I$. 

\begin{figure}
\begin{center}
\includegraphics[width=5cm]{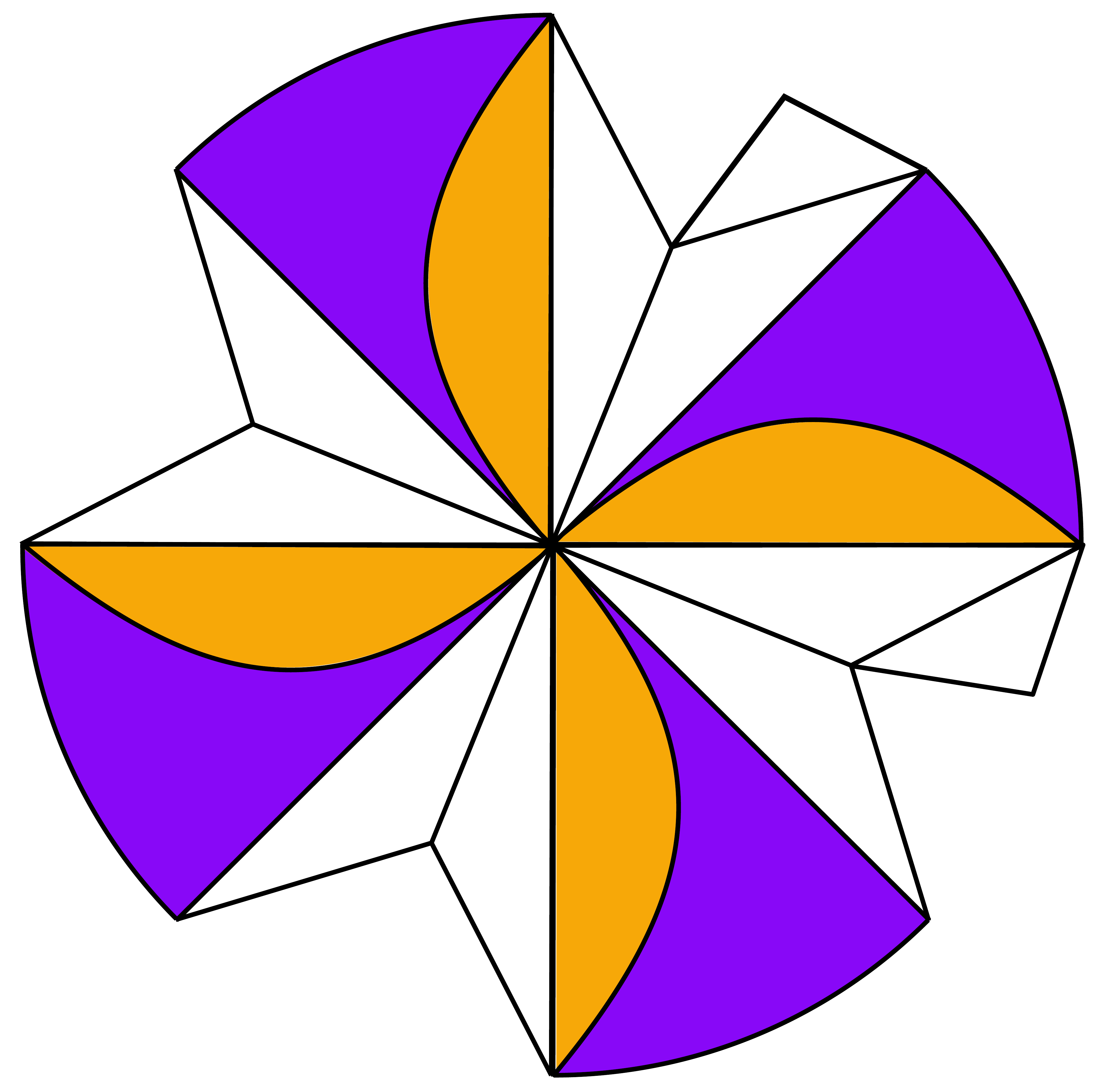}
\caption{\label{Comp2}The complex for the cohomological contextuality proof of the iffy scenario, in a different colouring. Orange: faces corresponding to the stabilizer relation Eq.~(\ref{SR2a}), purple: faces stemming from the stabilizer relation Eq.~(\ref{SR2b}),  white: faces invoking the recoupling relations Eq.~(\ref{esRel}).}
\end{center}
\end{figure}

We first consider the case of $N=4$ which is displayed in Fig.~\ref{Comp1}, and then the general case. Denote ${\cal{F}}:=\sum_i f_i$, i.e. ${\cal{F}}$ is the complete surface shown. It is easily verified that, after identifying the outer edges, $\partial {\cal{F}}=0$. Further, there are 9 faces in ${\cal{F}}$ on which $\beta_\Psi$ evaluates to 1, hence $\beta_\Psi({\cal{F}}) = 1\mod 2$. Both facts together imply that $[\beta_\Psi]\neq 0$, and hence the arrangement is contextual. $\Box$\medskip

We now turn to the general case of even $N$. If and only if $N$ is even, there is an even number of edges labeled by $X^{(3)}$ in the boundary of the disc (shown in Fig.~\ref{Comp1} for $N=4$). Hence $\partial {\cal{F}} = 0 \mod 2$ if and only if $N$ is even. We still need to establish $\beta_\Psi({\cal{F}}) = 1 \mod 2$. So let's count the number of faces $f$ with $\beta_\Psi(f)=1$. Such faces arise through the relation of Eq.~(\ref{StabRel2}), and there are $2N$ of them. Hence their contribution cancels mod 2. 

There is one more contribution to $\beta_\Psi({\cal{F}})$. For guidance, we look at Fig.~\ref{Comp1} and follow the red arrow in the counter-clockwise sense. The first observable we encounter that has non-trivial support only on qubit \#1 is $X_0^{(1)}$. The next such observable is $X_1^{(1)}$, then $X_2^{(1)}$, and so forth. Going around the disk, we increase the value of $k$ for such observables $X_k^{(1)}$ in increments of 1. Completing the circle, we arrive at $X_N^{(1)}$ which equals $-X_0^{(1)}$ by virtue of Eq.~(\ref{XkDef}). $X^{(1)}_0$ already is the label of the start-stop edge, and hence we obtain an additional factor of $-1$ (That is why, in Fig.~\ref{Comp1}, the color of the last face before completing the circle is white, $\beta_\Psi(f_\text{last})=0$). We have thus overcounted the contributions stemming from Eq.~(\ref{StabRel2}) by 1, which we now correct for. There are no other contributions, hence $\beta_\Psi({\cal{F}})=1$. 

Now assume the existence of a value assignment $s=(a_k,b_l,c_m)$, i.e., $\beta_\Psi=ds$. Then,
$$
1 = \beta_\Psi({\cal{F}}) = ds({\cal{F}}) = s(\partial_R {\cal{F}}) = s(0) = 0.
$$ 
Contradiction. Thus, no non-contextual value assignment exists. $\Box$\medskip

To conclude, let's compare the above proof for the iffy scenario with the original iffy proof. The ``iffiness" is gone. The algebraic structure Eq.~(\ref{IP}) underlying the iffy proof is replaced by a simpler one, namely a relative chain complex ${\cal{C}}(N)$ with 2-cocycle $\beta_\Psi(N)$ living in it ($N$ even). This is exactly the same structure as in the parity-based contextuality proofs without temporal order. 

We achieved this reduction to the prior case by introducing additional observables in the chain complex, namely $\{\epsilon_k, \sigma^+_k, \sigma^-_k\}$  as defined in Eq.~(\ref{EpSig}), to represent the temporal ordering. We propose this as a blueprint for a general method of constructing cohomological contextuality proofs describing temporally ordered measurement-based quantum computations. 

\section{Conclusion}\label{Concl}

In this paper, we have explained the contextuality--MBQC--cohomology triangle of Diagram~(\ref{Triangle}). Its upper corners, contextuality and measurement-based quantum computation, represent the phenomenology of interest; and the lower corner, cohomology, the mathematical method to describe it. The link between MBQC and contextuality is provided by Theorems~\ref{NLPCrel} and \ref{T1}, the link between contextuality and cohomology by Theorem~\ref{CPth2}, and the link between MBQC and cohomology by Theorem~\ref{ObeT} and Corollary~\ref{CohoCompCor}. Finally, in the center of the diagram sits the cocycle class $[\beta_\Psi]$, an element of the second cohomology group of the underlying chain complex. It contains the function computed in a given MBQC up to gauge equivalence, and the corresponding contextuality proof.

A limitation of the cohomological framework established to date is that it only applies to temporally flat MBQCs, which form a small subclass. Here we made a first step towards describing MBQCs and contextuality proofs  with temporal order in a cohomological fashion, by providing a cohomological contextuality proof in one concrete temporally ordered setting, the so-called ``iffy'' scenario \cite{Exa}.   Extending the cohomological formalism to all MBQCs with proper temporal order is a main subject of future research on the MBQC-contextuality connection.\medskip

\noindent
{\em{Acknowledgments.}} The author thanks the Yukawa Institute for Theoretical Physics Kyoto (YITP) for their hospitality. Part of this work was performed there. This work is supported by NSERC.

\section{Travel log}\label{TL}

As I learned over the years, the 8th Conference on Quantum Physics and Logic, held in Nijmegen, the Netherlands in November 2011, is remembered fondly by many participants; for all sorts of reasons. Here I'd like to describe my journey towards this conference, how I spiralled out of it, and my thoughts for the future.

My journey began in Munich in 2003, the final year of my PhD. Hans Briegel and I had discovered the one-way quantum computer, a scheme of measurement-based quantum computation (as it is now known) in 2000, and had answered the obvious first question---universality. Quite naturally, the universality proof was based on a mapping to the circuit model. But, besides proving the point, the mapping seemed inadequate in many ways. For example, the temporal order among the measurements in MBQC was different and more flat than the mapping would suggest: all Clifford gates can be implemented in the first round of measurement, before all other gates, irrespective of where they are located in the simulated circuit. This and similar observations prompted us to look for a description of MBQC outside the realm of circuit simulation, and, in the first place, for the basic structures upon which such a description could be built. 

There was, and is, no manual for how to approach this question. We are left to our own intuition and judgement. A structural element we focussed on early were the correlations among measurement outcomes that yield the computational result. Individually, the measurement outcomes in MBQC are completely random, and meaningful information can only be gleaned from certain correlations among them. What made the analysis of these correlations simultaneously difficult and interesting was their non-stabilizerness; i.e. the fact that the correlator observables are in general not mere tensor products of Pauli operators $X$, $Y$, $Z$.

Fault-tolerance seemed a path to make progress on these correlations. I figured that it could not be established for MBQC without understanding the structure of these correlations first. At the time, fault-tolerance with high error threshold was a problem with a price tag. In addition, when solved for MBQC we could sure learn something from the solution---a goldilocks problem. 

When first putting non-stabilizer quantum correlations on my map in early 2003, unbeknownst to me, someone in far away Moscow was finding out something about them: Sergey Bravyi. The next year we would be office mates at Caltech. 

Having arrived at Caltech in October 2003, it took about two years until, resting upon the scrap of two unsuccessful attempts, I established fault-tolerant universal MBQC with 3D cluster states \cite{FT1},\cite{FT2} (joint work with Jim Harrington and Kovid Goyal). Price tag fetched: the fault-tolerance threshold was high, and the whole construction elegant. 

And yet, one thing didn't completely fall into place---the learning-from-the-solution part. As noted above, I had stipulated that in order to establish fault-tolerance for MBQC, the structure of the non-stabilizer correlations would need to be understood first. It panned out differently. Those correlations did not need to be understood, and I hadn't understood them. This realization is one of three waypoints encountered at Caltech on my journey to Nijmegen. 

However, some correlations in MBQC---those which provide the error-correction capability for Clifford gates---could be understood very well. Namely, it turned out that those correlations have a cohomological underpinning. 3D cluster states can be described by a pair of three-dimensional chain complexes, related by Poincare duality. The measurement outcomes live on the respective faces, and are thus represented by 2-cochains $s$. The cluster state stabilizer implies that, in the absence of errors, the measurement record satisfies the constraint $s(\partial v)=0$, for all volumes $v$, and hence $s$ is a 2-cocycle. Furthermore, the output of the MBQC is given by evaluations $s(f)$, for non-trivial 2-cycles $f$. Fault-tolerance and computation on 3D cluster states is thus a matter of cohomology. This finding is the second Caltech waypoint.

In 2004, Sergey Bravyi and Alexei Kitaev developed ``magic state distillation'' \cite{BK}, an efficient and robust technique for implementing non-Clifford gates fault-tolerantly. It was eventually incorporated into fault-tolerant MBQC, but its main effect on me was a different one. Magic state distillation exploits non-Pauli quantum correlations to operate, as they are found, for example, in Reed-Muller quantum codes. Save the aspect of temporal order, these were precisely the quantum correlations I wanted to understand in the first place! 

A shortcut seemed to open:  What about using quantum Reed-Muller code states as computational resource states in MBQC---could toy computations exhibiting non-trivial correlations be constructed this way? I was eager to try, and settled on the following conditions for Reed-Muller toy MBQCs: (i) The classical side processing relations Eq.~(\ref{CCR}) have to be obeyed; in particular, the input values form a vector space, as required by Eq.~(\ref{CCR_in}). (ii) The outcome is deterministic for every admissible value of input, and (iii) the MBQC is non-Clifford. Further, the criterion for an ``interesting'' computation was that it computed a non-linear Boolean function. Quite a low bar, but justified as it exceeds what the classical side processing permits by itself.

\begin{figure}
\begin{center}
\includegraphics[width=16cm]{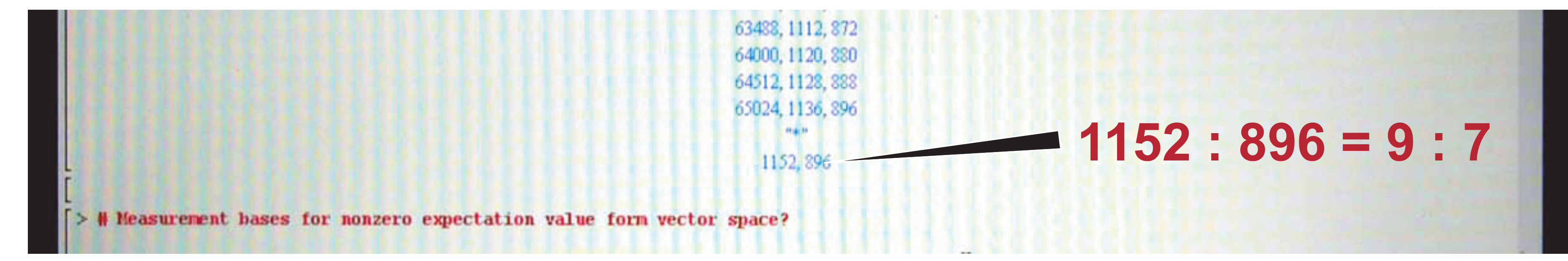}
\caption{\label{Maple}Numerical experiment on toy MBQCs using Reed-Muller quantum code states as computational resources. Shown is the output for the example based on a 31-qubit punctured Reed-Muller code. All tests worked out---the Boolean function computed was total and non-linear.}
\end{center}
\end{figure}

Armed with those criteria, I got my laptop running. I started with the 15 qubit punctured Reed-Muller quantum code, and it didn't work. So I went on to the 31 qubit punctured Reed Muller code, which, given the next came at 63, I knew was the largest I could handle. I held my breath. There was deterministic output on 2048 inputs---power of 2, good sign. The output was imbalanced, hence the computed function non-linear. A final check remained to be made: did the inputs form a vector space, as required by Eq.~(\ref{CCR_in})? That worked out too! I was over the moon. 

Sometime in the subsequent months, while finalizing the fault-tolerance work, it must have trickled in that to be excited about such toy quantum computations needed a very particular taste or preparation. They didn't achieve anything of real computational value. At any rate, the finding of these Reed-Muller toy MBQCs is my third waypoint at Caltech.

In 2008, after I had moved to the University of British Columbia by way of the Perimeter Institute (PI), at a workshop at PI I heard Dan Browne speak about similar toy MBQCs. In a work of Janet Anders and him \cite{AB}, they considered MBQCs on a Greenberger-Horne-Zeilinger state, satisfying the above conditions (i) and (ii). Not enforcing condition (iii) (non-Cliffordness) allowed them to get by with 3 qubits rather than 31. But much, much more importantly, they managed to relate their 3 qubit-MBQC to something known and valued in the world of Physics: Mermin's star. Thus the MBQC--contextuality link saw the light of day. Learning of this result I was ready to go to QPL 2011, although the conference was still 3\,1/2 years ahead.\medskip

Finally, being at QPL 2011 in Nijmegen, what made the day for me was a talk by Samson Abramsky, Shane Mansfield and Rui Barbosa on ``The Cohomology of Non-Locality and Contextuality''. It had taken me quite a bit of effort to make it to the conference---teaching had to be rescheduled and so on. But I boarded the return plane in Amsterdam with a swagger: very, very worth the trouble. Although, honestly, in actual terms I had not learned all that much. I had understood precisely one slide of Shane Mansfield's  presentation, and that was the title slide. What my journey through Caltech and PI had prepared me for was to see significance in the words ``contextuality'' and ``cohomology'' appearing side by side. I also somehow managed to not be completely bypassed by Mansfield's cohomological explanation of the GHZ scenario, at least in so far as I noted the argument's existence. Of course I tried to chase down Mansfield and Barbosa after their talk, but they seemed quite busy answering other calls.\medskip

For me, the upshot of Nijmegen was that a cohomological theory of MBQC was in range, making sense of all the known toy examples and hopefully beyond. To get started, all I needed to do was to get to grip with the Abramsky--Mansfield--Barbosa paper  \cite{A2}, which finally happened in the Spring of 2012. Then it turned out that their cohomological explanation of the GHZ example did not quite provide the desired connection to MBQC. The latter required a cohomological interpretation of precisely Mermin's argument for the GHZ-scenario, not merely a cohomological explanation of that scenario. And so, with my collaborators Cihan Okay, Stephen Bartlett, Sam Roberts and Emily Tyhurst, we set out to define our own cohomological framework. I do not need to describe the ensuing work here, since I already did in the previous sections.\medskip

This brings me to my thoughts for the future. Regarding measurement-based quantum computation, the recent investigations into its structure---contextuality as we discussed it here,   computational phases of matter \cite{screen}--\cite{Archi} and temporal order \cite{CompMod}, \cite{Gflow}, \cite{TO_sym}---have to day remained separate. And yet they share a common ingredient at their cores: symmetry. I'm confident that these investigations will be unified into a single framework in the coming years, and that something new will spring from it.

To think about the future of our field more broadly, let's take a really long run-up and zoom right into the year 1842.  Ada, Countess of Lovelace and assistant to the British computing pioneer Charles Babbage, had just invented the notion of the computer program. Also, at a time when everybody around her saw the future of computation in calculating trajectories of cannon balls, she had the fundamental insight that not only numbers can be processed by computers, but rather symbolic information of any kind---musical notes, images, text \cite{Innovators}. Her insight lives on today in digital radio and television, the internet, Maple, the Google search engine, and countless other inventions of the information age. 

But, quantum computation extends beyond this line of thought. Quantum information is not ``symbolic''. Due to the irreversibility of quantum measurement, it cannot be perceived by looking at it. And with the limits of the reigning paradigm exposed, a new era of computation can begin---at least in the skunkworks. On the theory side of it, whether one is thinking about measurement-based quantum computation or the circuit model, essentially everything boils down to one thing: quantum algorithms. Towering achievements such as Shor's factoring not\-with\-standing, we seem to have difficulty inventing new quantum algorithms, and it's a matter of intuition. 

\begin{center}
What would Ada's insight be today?
\end{center}

\end{document}